\documentclass[a4paper,twocolumn,10pt]{article}

\pdfoutput=1

\usepackage{graphicx}
\usepackage{bm}
\usepackage{dcolumn}
\usepackage{setspace}
\usepackage{abstract}
\usepackage{multirow}

\usepackage[utf8]{inputenc}
\usepackage{lineno,hyperref}

\begin{document}
\title{ \bf Deformation behaviour of body centered cubic Fe nanowires under
tensile and compressive loading}
\date{}
\author{\small G. Sainath\footnote{email : sg@igcar.gov.in},  B.K. Choudhary and T.
Jayakumar \\ 
\small Metallurgy and Materials Group \\ 
\small Indira Gandhi Centre for Atomic Research, Kalpakkam \\
\small Tamilnadu-603102, India }
\twocolumn[
\maketitle

\begin{onecolabstract}
 
Molecular Dynamics (MD) simulations have been carried out to investigate the
deformation behaviour of $<$110$>/\{111\}$ body centered cubic (BCC) Fe 
nanowires under tensile and compressive loading. An embedded atom method (EAM) 
potential was used to describe the interatomic interactions. The simulations 
were carried out at 10 K with a constant strain rate of $1\times10^8 s^{-1}$. 
The results indicate a significant differences in deformation mechanisms under 
tensile and compressive loading. Under tensile loading, the deformation occurs 
by the slip of full dislocations, While under compressive loading twinning was 
observed as the dominant mode of deformation. The tension-compression asymmetry
in deformation mechanisms of BCC Fe nanowires is attributed to
twinning-antitwinning asymmetry of 1/6$<$111$>$ partial dislocation on \{112\}
planes. We further explain the mechanism of dislocation pile up in tensile
loading and twin growth in compressive loading. \\
 
\noindent {\bf Keywords: } Molecular Dynamics; Nanowires; Tension; Compression; 
Slip; Twinning.

\noindent {\bf PACS : }  83.10.Rs, 81.07.Gf, 62.20.M-, 61.72.Hh, 61.72.Mm \\


\end{onecolabstract}

]
\renewcommand{\thefootnote}{\fnsymbol{footnote}} \footnotetext{* email :
sg@igcar.gov.in}

\section{Introduction}

Understanding the deformation behavior of nanscale materials has become a major 
interest for material scientists due to their potential applications in
nano/micro electro-mechanical systems (NEMS/MEMS). BCC Fe 
nanowires with superior magnetic properties are useful in medical sensors, 
data-storage media, spin electronics and other memory devices \cite{Zhang,
American}. The complexities involved in performing experiments at nanoscale 
preclude the conventional testing methods and lend towards
theoretical/computational tools. With the rapid progress of computational 
capability and the availability of reliable inter-atomic potentials, MD 
simulations have become a major tool to probe the mechanical behaviour of 
nanoscale materials. Moreover, MD simulations provide the atomistic detail 
of the deformation and damage mechanisms. In this work, atomistic simulations 
were utilized to investigate the mechanical behaviour of BCC Fe nanowires.

Many experimental and simulation studies have shown that the plastic deformation
in nanoscale materials occurs through deformation twinning, dislocation slip
and phase transformation \cite{Park,Progress,JMC}. The competition between
these mechanisms mainly depends on crystallographic orientation, size and mode 
of loading (tension/compression). Particularly, with respect to mode of loading 
often an asymmetry is observed in deformation mechanisms of metallic nanowires. 
For example, the $<$100$>$ oriented Au and Cu nanowires deforms by full
dislocation slip under tensile loading, while partial dislocation slip is
observed under compressive loading \cite{Park}. Similarly, the $<$110$>$
oriented Au and Cu nanowires deforms by partial slip under tensile loading, 
while full slip is observed under compressive loading \cite{Park}. These
results shows that the asymmetry in deformation mechanisms is orientation 
dependent. In face centered cubic (FCC) metals, slip occurs on \{111\} planes 
in $<$110$>$ directions and the planar nature of the dislocation cores make 
the slip to follow the schmid's law and the asymmetry in deformation mechanisms 
can be attributed to different schmid factors for leading and trailing 
partial dislocations \cite{Park,JMC}. In general, most of the studies in 
literature are mainly focused upon FCC metallic nanowires and comparatively 
few studies exist on their BCC counterparts \cite{Kim,Healy}. In BCC metals, 
the slip occurs in $<$111$>$ direction along \{110\}, \{112\} and \{123\} 
planes and the non-planar nature of screw dislocations core makes the slip 
to violate the schmid's law. In Mo nanopillars, Kim et al.\cite{Kim} studied the
crystallographic orientation dependence of tension-compression asymmetry 
and shown that compressive flow stresses are higher than tensile ones in 
$<$100$>$ orientation, and vice versa for $<$110$>$ orientation. Healy and 
Ackland \cite{Healy} had shown that the $<$100$>$ oriented BCC Fe nanopillars 
deforms by twinning mechanism under tensile loading, while full dislocation 
slip is observed under compressive loading. However, it is not clear how the 
deformation mechanisms change with respect to mode of loading in other 
orientations, for example in $<$110$>$ and $<$111$>$ orientations. The 
present investigation is aimed to understand the deformation mechanisms under 
tensile and compressive loading of $<$110$>$/\{111\} BCC Fe nanowires using 
MD simulations.

\section{Simulation details}
Molecular Dynamics (MD) simulations have been carried out in Large scale 
Atomic/Molecular Massively Parallel Simulator (LAMMPS) package \cite{LAMMPS} 
employing an embedded atom method (EAM) potential for BCC Fe given by Mendelev 
and co-workers \cite{Mendelev}. This potential was based on the framework of 
a Finnis-Sinclair (FS) type many body potential. The potential was fitted to 
the properties obtained using first-principles calculations in a model liquid 
configuration and also to other experimentally obtained material properties 
of BCC Fe.

Single crystal BCC Fe nanowire oriented in $<$110$>$ axial direction with
\{111\} as side surfaces were considered for this study. The nanowire had 
dimensions of $8.5\times8.5\times17$ $nm^3$ ($30a\times30a\times60a$, where 
$a = 2.855$ $A^0$ is lattice parameter of BCC Fe). The simulation box contains
about 110000 Fe atoms arranged in BCC lattice. The nanowire length was set 
as twice the cross- section width (d) (aspect ratio 2:1 ). The fixed aspect 
ratio ensures, that the effects that are dependent on aspect ratio were 
eliminated. Periodic boundary conditions were chosen only along the wire 
axis, while the other directions were kept free in order to mimic an 
infinitely long nanowire. After initial construction of the nanowire, 
energy minimization was performed by conjugate gradient (CG) method to 
obtain a stable structure. The stable structure thus obtained is thermally 
equilibrated to a temperature of 10 K in canonical ensemble ( constant NVT ) 
by choosing the initial velocities of atoms randomly from a finite 
temperature Maxwell distribution. Velocity verlet algorithm was used to 
integrate the equation of motion with a time step of 5 fs. 

Upon completion of the equilibration process, the tensile and compressive
loadings were carried out at constant strain rate of $1\times10^8 s^{-1}$. The
loading was applied along the nanowire axis ($<$110$>$). The strain rates
considered here are very much higher than typical experimental strain rates 
due to the inherent timescale limitations from MD simulations. During the
loading, the atomic system was allowed to deform naturally at constant strain 
rate, without imposing any stress constraints in other two directions. The 
strain is defined as $L-L_{0}/L_{0}$, where L is the current length of the 
nanowire and $L_0$ is the length just after the relaxation. The stress is 
calculated from the Virial expression of stress \cite{Virial}, which is 
equivalent to a Cauchy's stress in an average sense. Atom-Eye package 
\cite{AtomEye} is used for visualization of atomic snapshots with 
centro-symmetry parameter (CSP) coloring \cite{CSP}. The burger vector of
a dislocations is determined by the Dislocation Extraction Algorithm (DXA) 
developed by Stukowski \cite{DXA}. 

 
 \begin{figure}
  \centering
 \includegraphics[width=7cm]{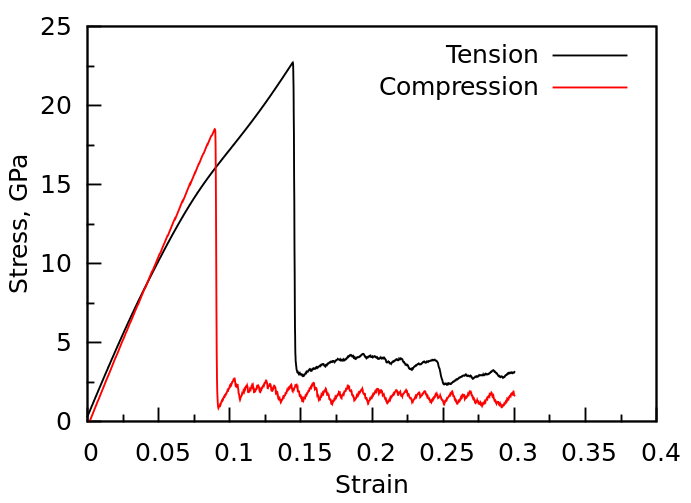}
 \caption {\small The stress-strain curves of $<$110$>$/\{111\} BCC Fe nanowires
at 10 K under tensile and 
 compressive loading}
 \label{stress-straincurve}
 \end{figure}

\section{Results}
Fig.\ref{stress-straincurve} shows the stress-strain curves of
$<$110$>$/\{111\} BCC Fe nanowires under tensile and compressive loading upto
30\% of strain at 10 K. In both loading conditions, initially the nanowires
undergo an elastic deformation up to a peak stress, which can be designated as
an yield strength of a defect-free nanowire. The Young’s moduli obtained from
the stress-strain curves was 200 GPa and it show no striking difference between
the two loading conditions. Similarly, the magnitudes of the yield stress in
tensile and compressive loadings were found to be 23 GPa and 18.5 GPa,
respectively. The strength values are in close agreement with those obtained by
ab-initio calculations for BCC Fe \cite{Chinese}. Following the peak, the flow
stress drops abruptly to low values of stress, indicating the occurrence of
yielding in the nanowires. It was observed that, the yielding under compressive 
loading occurred much earlier than in tensile loading. Further, it can be seen 
that, the magnitude of yield stress, flow stress and yield strain were higher in 
tensile than in compressive loading. 


\begin{figure}
\centering
 \includegraphics[width=7.5cm]{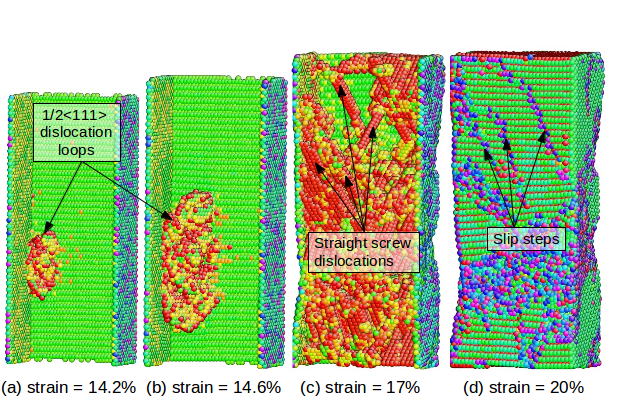}
 \caption {\small The atomic snapshots showing the deformation behaviour of
$<$110$>$/\{111\} BCC Fe nanowires under tensile  loading. (a) The nucleation 
(b) expansion of dislocation loops (c) accumalation of straight screw 
 dislocations and (d) the slip steps formed due to dislocation escape. The 
 atoms are colored according to the centro-symmetry parameter (CSP) \cite{CSP}. 
 The perfect BCC atoms and the front surfaces are removed for clarity in (a-c).}
 \label{deformation-tensile}
 \end{figure}

\begin{figure}
\centering
 \includegraphics[width=7cm]{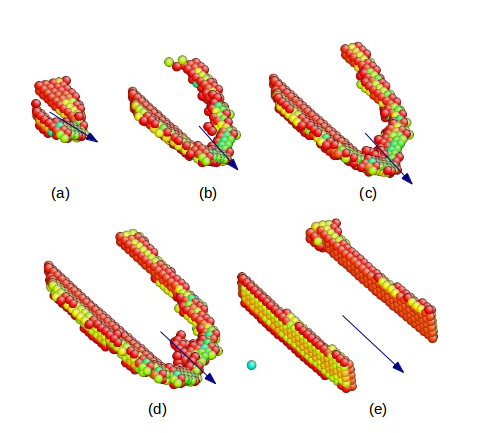}
 \caption {\small The typical mechanisms of straight screw dislocations pile-up
in $<$110$>$/\{111\} BCC Fe nanowires under tensile loading. The atoms on 
dislocation cores are only shown. The blue arrow represents the burger vector 
direction}
 \label{straight-screws}
 \end{figure}

In order to understand the deformation behaviour under tensile and compressive
loading, the evolution of atomic configurations at various strain levels were 
analysed using Atom-Eye package \cite{AtomEye} with centro-symmetry parameter 
(CSP) coloring \cite{CSP}. Fig. \ref{deformation-tensile} shows the deformation 
behaviour under tensile loading of $<$110$>$/\{111\} BCC Fe nanowires at 10 K.
It can be seen that the nanowire yields by the nucleation of collective 
dislocation loops from a single source at the corner of the nanowire 
(Fig. \ref{deformation-tensile}(a)). The DXA analysis indicates that, the 
dislocation loops had a screw character with burger vector of 1/2$<$111$>$ 
(full dislocations). The yielding in the nanowire causes an abrupt drop in flow 
stress from a peak to a low level of stress. Previous studies on single crystal 
Fe-3\%Si had shown that, once the stress reaches close to the theoretical 
strength, several dislocations loops might nucleate from the same source
\cite{Indent}. Since, there are no obstacles present in the nanowire, once 
nucleated avalanche of dislocations loops easily expand in multiple directions 
as shown in Fig. \ref{deformation-tensile}(b). With increasing deformation, the
accumulation of large number of screw dislocations can be seen in Fig.
\ref{deformation-tensile}(c). The pile-up of straight screw dislocations in 
BCC metals has been discussed previously by Groger and Vitek \cite{Groger} 
with a mesoscopic model and Kaufmann et al. \cite{Kaufmann}, using kinetic 
pile-up model. Experimentally Kim et al. \cite{Kim}, had shown that, when 
mixed dislocation loops are present in the sample during annealing, its 
edge components move and annihilate at a faster rate than the screw 
components. As a result, the screw dislocations likely remain in the
metal, post annealing. The typical mechanism of screw dislocation 
accumulation in BCC Fe nanowires is shown in shown in Fig.
\ref{straight-screws}. The dislocation loops initially emitted from the 
nanowire corner had a mixed character (Fig. \ref{straight-screws}(a)). With
increasing strain, the expansion of dislocation loops associated with the
decrease in the curvature can be seen in Fig. \ref{straight-screws}(b-c). Since,
edge dislocations in BCC metals have higher mobility than the screw
dislocations, the edge component of mixed dislocations can easily reaches the
surface (Fig. \ref{straight-screws} (d) and (e)) and annihilates, leaving a well
defined slip steps on the nanowire surface and many of such slip steps can be
seen in Fig. \ref{deformation-tensile}(d). As a result of edge dislocation
escape, only screw component of a dislocation loop remains in the nanowire 
(Fig. \ref{straight-screws}(e)).

 
\begin{figure}
\centering
 \includegraphics[width=7.5cm]{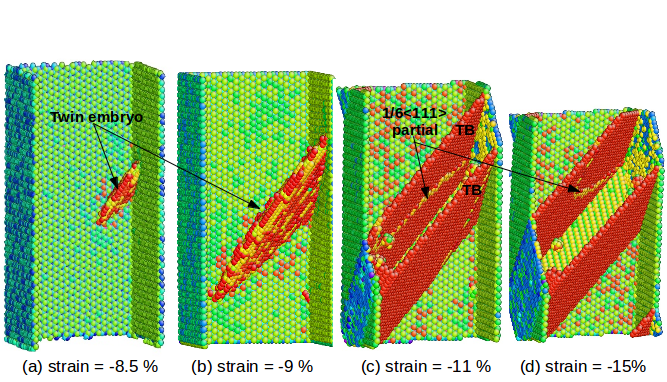}
 \caption {\small The atomic snapshots showing the deformation of
$<$110$>$/\{111\} BCC Fe nanowires  under compressive loading. (a) The
nucleation (b) growth of twin embryo (c) forming a full twin plate and 
(d) twin growth. The atoms are coloured according to the centro-symmetry 
parameter (CSP) \cite{CSP}. The perfect BCC atoms and front surfaces are 
removed for clarity.}
 \label{deformation-compression}
 \end{figure}

Fig. \ref{deformation-compression} shows the atomic configurations at various 
stages of deformation during the compressive loading of $<$110$>$/\{111\} BCC 
Fe nanowires. It can be seen that the yielding under compressive loading occurs 
by the nucleation of a twin embryo from the corners of the nanowire 
(Fig. \ref{deformation-compression}(a)), causing an abrupt drop in flow stress. 
The initially nucleated twin embryo propagate towards the other surface 
(Fig. \ref{deformation-compression}(b)). The twin embryo consists of 
1/6$<$111$>$ partial dislocations along with 'twin' like stacking faults on 
\{112\} planes (Fig. \ref{deformation-compression}(a-b)). With increasing
strain, the leading edge of a twin embryo reaches the opposite surface and
becomes full twin enclosed by twin boundaries on \{112\} planes as shown in 
Fig. \ref{deformation-compression}(c). With further increase in deformation,
the twin grows by the motion of twin boundary along the nanowire axis 
(Fig. \ref{deformation-compression}(c-d)). The twin growth or twin boundary 
motion along the nanowire axis occurs by repeated initiation and glide of 
1/6$<$111$>$ twinning partial dislocations on adjacent \{112\} planes. It can
be seen that, the high stress is required for the twin nucleation, but
comparatively low stress is sufficient for twin growth or twin boundary motion.
This is because the twin nucleation involves the creation of new fault area,
where as the twin growth will not create any new fault area and therefore low
energy is sufficient.  

 
\begin{figure}
\centering
 \includegraphics[width=7cm]{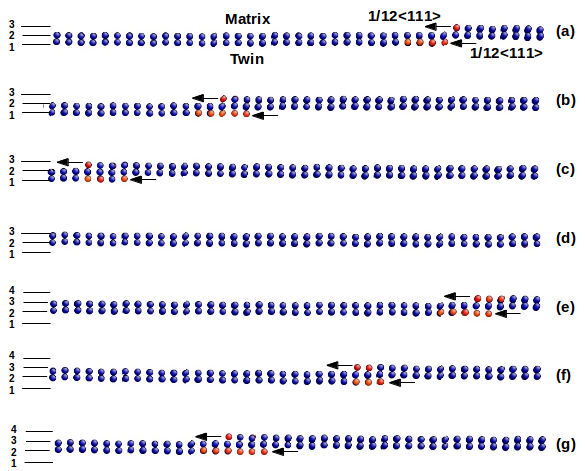}
 \caption {\small The process of twin growth in $<$110$>$/\{111\} BCC Fe
nanowires under compressive loading. The twin growth proceeds by repeated 
initiation and glide of 1/6$<$111$>$ (= 1/12$<$111$>$ + 1/12$<$111$>$ ) partial 
dislocations on adjacent \{112\} planes. The blue atoms represents \{112\} twin
boundary and arrow marks on twin boundary show the 1/12$<$111$>$ partial 
dislocations. For clarity the twin boundary layers are marked with numerics 
1,2,3 and 4}
 \label{TB-structure}
 \end{figure}

Fig. \ref{TB-structure} shows the process of twin growth under the compressive
loading of $<$110$>$/\{111\} BCC Fe nanowires. As shown in the Fig.
\ref{TB-structure}, the twin boundary is of displaced type \cite{Yamaguchi},
consisting of two 1/12$<$111$>$ dislocations lying in the adjacent \{112\}
planes, marked 1 and 3 in Fig. \ref{TB-structure}(a). The 1/6$<$111$>$ (= 
1/12$<$111$>$ + 1/12$<$111$>$ ) partial dislocations propagates further on
\{112\} twin boundaries as shown in Fig. \ref{TB-structure}(b-c) and annihilates
at the opposite surface (Fig. \ref{TB-structure}(d)). The splitting of
1/6$<$111$>$ partial dislocation in BCC Fe is in agreement with the observation
made by Yamaguchi et al. \cite{Yamaguchi}. Again the next 1/6$<$111$>$ twinning
partial dislocation nucleates next to the first one (Fig.
\ref{TB-structure}(e)), glides on the adjacent \{112\} plane (Fig.
\ref{TB-structure}(f-g)) and again annihilates at the opposite surface. By
repeating this process, the twin grows along the nanowire axis. Each nucleation
and annihilation of twinning partial dislocation removes one inter-planer
spacing from the parent lattice and adds to twin lattice ( In Fig.
\ref{TB-structure}(a-d), the layer 3 is removed from the parent lattice and 
correspondingly the layer 1 is added to twinned lattice ). This mechanism 
results the twin plate to thicken by a layer by layer growth process. The twin
growth during the deformation results in a constant plateau observed in flow
stress. 

 
\section{Discussion}
The stress-strain behaviour of $<$110$>$/\{111\} BCC Fe nanowires had shown that
the yield stress under tensile loading is higher than that of compressive
loading. It indicates that the yield strength exhibits tension-compression 
asymmetry. Similar to the present results, many atomistic simulations studies on
FCC and BCC metallic nanowires have shown an asymmetry in yield stress 
\cite{Kim,Tschopp,Gold}. In FCC metallic nanowires it has been attributed to
the presence of intrinsic stress in the nanowire caused by surface stress
\cite{Gold}. The present results show that, the tension-compression asymmetry in
BCC Fe nanowires is mainly due to distinct differences in yielding mechanisms.
Since the yield stress is associated with defect nucleation, the distinct
yielding mechanisms leads to a tension-compression asymmetry in yield stress. 
Under tensile loading, the yielding occurs by the nucleation of high energy full
dislocations (Fig. \ref{deformation-tensile}(a)), whereas in compression, it 
occurs by the nucleation of relatively low energy partial dislocations which 
causes twinning (Fig. \ref{deformation-compression}(a)). Another factor
influencing the tension-compression asymmetry of BCC Fe nanowires is related 
to the intrinsic properties of screw dislocations in BCC metals
\cite{Kim,Vitek}. The core of 1/2$<$111$>$ screw dislocations in BCC metals
spreads into several non-parallel planes of $<$111$>$ zone \cite{Vitek}. As a
result, the glide of a screw dislocation depends on the shear stress in slip 
direction along with the stress perpendicular to the slip direction, known as 
non-glide stress. The non glide stress is positive for tension, facilitating 
the easy glide and analogously it is negative for compression, making the
dislocation glide difficult \cite{Kim-2}. 

\begin{table}
\caption{Deformation mechanisms} \
\centering
\scalebox{0.85}{
\begin{tabular}{lll} \hline
\multirow {2}{*}{Loading type} & Loading axis\\
 & $<$110$>$ & $<$100$>$ \\ \hline  
Tension & Dislocation slip{*} & Twinning \cite{Healy}\\ 
Compression & Twinning {*} & Dislocation slip \cite{Healy} \\ \hline
{*} \small present study
\end{tabular} }
\label{Table}
\end{table}

Table \ref{Table} summarizes the deformation mechanisms observed in $<$110$>$/\{111\} BCC
Fe nanowires to that observed in $<$100$>$/\{110\} nanowires by Healy and
Ackland \cite{Healy}. Healy and Ackland \cite{Healy} observed twinning under
tensile loading and dislocation slip under compressive loading of
$<$100$>$/\{110\} BCC Fe nanowires. It can be seen that the tension-compression 
asymmetry in deformation mechanisms of $<$110$>$/\{111\} BCC Fe nanowires is 
distinctly different than what has been observed in $<$100$>$/\{110\} BCC Fe 
nanowires. This is mainly because, the $<$100$>$ and $<$110$>$ orientations 
show a opposite twinning-antitwinning slip on \{112\} planes \cite{Kim}. Under 
compressive loading of $<$110$>$/\{111\} BCC Fe nanowires, the 1/6$<$111$>$
partial dislocation glides in twinning sense leading to the deformation by 
twinning. While in tension, it glides in antitwinning sense (opposite
direction), making the dislocation glide more difficult on \{112\} planes. As a
result, the dislocation may prefer to glide on secondary slip planes such as
\{110\} or \{123\} leading to the deformation through slip in anti-twinning
direction. This mechanisms is reversed for $<$100$>$/\{110\} BCC Fe nanowires.
Therefore, the observed tension-compression asymmetry in deformation behaviour of
BCC Fe could be due the twinning-antitwinning asymmetry of 1/6$<$111$>$ partial 
dislocation on \{112\} planes. 

\section{Conclusion}

In summary, molecular dynamics simulations on $<$110$>$/\{111\} BCC Fe
nanowires show that the deformation mechanisms vary with mode of loading. 
Under tensile loading, the deformation occurs by the full dislocations 
and the pile up of straight screw dislocations is observed. Whereas under 
compressive loading, the twinning was observed as predominant mode of 
deformation and the growth of twin occurred through the repeated initiation 
and glide of 1/6$<$111$>$ (= 1/12$<$111$>$ + 1/12$<$111$>$ ) partial 
dislocations on adjacent \{112\} planes. The twin boundary is found to be 
of displaced type, consisting of two 1/12$<$111$>$ dislocations lying in 
the adjacent \{112\} planes. Due to distinct yielding mechanisms, the 
yield strength exhibited tension-compression asymmetry. The observed 
tension-compression asymmetry in deformation behaviour of BCC Fe nanowires 
is attributed the twinning-antitwinning asymmetry of 1/6$<$111$>$ partial 
slip on \{112\} planes.

\end{document}